\documentclass[prl,twocolumn,floatfix,%
showpacs,preprintnumbers,amsmath,amssymb]{revtex4}
\usepackage{graphicx}
\def\half{{\textstyle \frac{1}{2}}}
\def\lf{\left} \def\ri{\right}
\def\sk{S} \def\sw{S_{\rm sw}}
\def\asy{a_{\rm sym}} \def\asya{a_{\rm sym}(A)}
\def\csy{c_{\rm sym}} \def\Ksy{K_{\rm sym}}
\def\beq{\begin{equation}} \def\eeq{\end{equation}}

\begin{document}
\title{Nuclear symmetry energy probed by neutron skin thickness of nuclei}
\author{M. Centelles$^1$} \author{X. Roca-Maza$^1$} 
\author{X. Vi\~nas$^1$} \author{M. Warda$^{1,2}$}
\affiliation{$^1$Departament d'Estructura i Constituents de la Mat\`eria
and Institut de Ci\`encies del Cosmos, Facultat de F\'{\i}sica, Universitat de Barcelona,
Diagonal {\sl 647}, {\sl 08028} Barcelona, Spain\\
$^2$Katedra Fizyki Teoretycznej, Uniwersytet Marii Curie--Sk\l odowskiej, 
ul.\ Radziszewskiego {\sl 10}, {\sl 20-031} Lublin, Poland}

\begin{abstract} 
We describe a relation between the symmetry energy coefficients
$\csy(\rho)$ of nuclear matter and $\asya$ of finite nuclei that
accommodates other correlations of nuclear properties with the
low-density behavior of $\csy(\rho)$. Here we take advantage of this
relation to explore the prospects for constraining $\csy(\rho)$ of
systematic measurements of neutron skin sizes across the mass table,
using as example present data from antiprotonic atoms. The found
constraints from neutron skins are in harmony with the recent
determinations from reactions and giant resonances.
\end{abstract} 
\date{\today}
\maketitle

A wealth of measured data on densities, masses and collective
excitations of nuclei has allowed to resolve basic features of the
equation of state (EOS) of nuclear matter, like the density
$\rho_0\approx0.16$ fm$^{-3}$, energy per particle $a_v\!\approx\!-16$
MeV, and incompressibility $K_v\approx230$ MeV \cite{pie07} at
saturation. However, the symmetry properties of the EOS due to
differing neutron and proton numbers remain more elusive to date. The
quintessential paradigm is the density dependence of the symmetry
energy \cite{pie07,bro00,hor01,ste05,bar05,li05,li08,fam06,she07,lat07}. 
The accurate characterization of this property entails profound
consequences in studying the neutron distribution in stable and exotic
nuclei and neutron-rich matter \cite{bro00,hor01,ste05}. It impacts on
heavy ion reactions \cite{bar05,li05,li08,fam06,she07}, nuclear
astrophysics \cite{hor01,ste05,lat07}, and on diverse areas such as
tests of the Standard Model via atomic parity violation~\cite{sil05}.

The general expression $e(\rho,\delta)= e(\rho,0) + \csy(\rho)\delta^2
+ {\cal O}(\delta^4)$ for the energy per particle of nuclear matter of
density $\rho=\rho_n+\rho_p$ and asymmetry
$\delta=(\rho_n-\rho_p)/\rho$ defines the symmetry energy coefficient
$\csy(\rho)$ of a nuclear EOS. It is customary and insightful to
characterize the behavior of an EOS around the saturation density
$\rho_0$ in terms of a few bulk parameters, like $e(\rho,0)\simeq a_v
+ \half K_v\epsilon^2$ and $\csy(\rho)\simeq J - L\epsilon +
\half\Ksy\epsilon^2$ where $\epsilon=(\rho_0-\rho)/(3\rho_0)$
\cite{bar05,li05,li08,mye69}. The value of $J= \csy(\rho_0)$ is
acknowledged to be about 32 MeV. The values of
$L=3\rho\partial\csy(\rho)/ \partial\rho |_{\rho_0}$ and
$\Ksy=9\rho^2\partial^2 \csy(\rho)/ \partial\rho^2 |_{\rho_0}$ govern
the density dependence of $\csy$ around $\rho_0$. They are less
certain and the predictions vary largely among nuclear theories, see
e.g.\ Ref.\ \cite{li08} for a review.

In experiment, recent research in intermediate-energy heavy ion
collisions (HIC) is consistent with a dependence
$\csy(\rho)=\csy(\rho_0) (\rho/\rho_0)^\gamma$ at $\rho<\rho_0$
\cite{li05,li08,she07,fam06}. Isospin diffusion predicts
$\gamma=0.7$--1.05 ($L=88\pm25$ MeV) \cite{li05,li08}, isoscaling
favors $\gamma=0.69$ ($L\sim65$ MeV) \cite{she07}, and a value closer
to 0.55 ($L\sim55$ MeV) is inferred from nucleon emission ratios
\cite{fam06}. Nuclear resonances are another hopeful tool to calibrate
$\csy(\rho)$ below $\rho_0$~\cite{gar07,kli07,tri08,lia08}. Indeed,
the giant dipole resonance (GDR) of $^{208}$Pb analyzed with Skyrme
forces suggests a constraint $\csy(0.1{\rm\,fm}^{-3})=23.3$--24.9 MeV
\cite{tri08}, implying $\gamma\sim0.5$--0.65. Note that the
Thomas-Fermi model fitted very precisely to binding energies of 1654
nuclei \cite{TF94} predicts an EOS that yields $\gamma=0.51$. With the
caveat that the connection of experiments to the EOS often is not at
all trivial \cite{li05,li08,she07,fam06,gar07}, it is important to
seek further clues from the above and other isospin-sensitive signals,
such as the neutron skin thickness $\sk=R_n-R_p$ of nuclei (difference
of neutron and proton rms radii). Because $\sk$ of heavy nuclei
correlates linearly with the slope $L$ of $\csy$ in mean field
theories of nuclear structure
\cite{bro00,hor01,bar05,li05,li08,ste05,fur02,dan03}, these studies
have far-reaching implications for nuclear theory.

In this work we show that $\csy(\rho)$ of the EOS equals at
$\rho\approx0.1{\rm\,fm}^{-3}$ the value of the symmetry energy
coefficient $\asya$ of heavy {\em finite} nuclei, universally in mean
field theories. The observed correlations of $\sk$
\cite{bro00,hor01,bar05,li05,li08,ste05} and of the excitation energy
of the GDR \cite{tri08} with the density dependence of $\csy$ can be
deduced naturally from this relation. We resort to the nuclear
droplet model (DM) \cite{mye69} to work out the analytical formulas.
The result derived for $\sk$ is applied to investigate limits to the
slope and curvature of $\csy$ from neutron skins measured for 26
stable nuclei, from $^{40}$Ca to $^{238}$U, in antiprotonic atoms
\cite{trz01}. A main point is ascertaining how far {\it uniformly}
measured neutron skins {\it over} the periodic table may help
constrain the density dependence of $\csy$. We provide first evidence
that the constraints from skins are in consonance with the recent
observations from reactions and giant resonances, though the probed
densities and energies are not necessarily the same.

The symmetry energy coefficient $\asya$ of finite nuclei is smaller
than the bulk value $J$. Given a nuclear force, the DM allows one to
extract $\asya$ as \cite{mye69,bra85}
\beq\label{eq4}
\asya= \frac{J}{1+x_A}, 
\quad \mbox{with }\,x_A= \frac{9J}{4Q}A^{-1/3} .
\eeq
The so-called surface stiffness $Q$ measures the resistance of the
nucleus against separation of neutrons from protons to form a neutron
skin. One can obtain $Q$ of nuclear forces by asymmetric semi-infinite
nuclear matter (ASINM) calculations \cite{mye69,bra85,cen98}. The
contribution of $\asya$ to the nucleus energy is $\asya\,(I+x_A I_C)^2
A$, where $I=(N-Z)/A$ and $I_{\rm C}= e^2 Z/(20J R)$ is due to
Coulomb. One has $R=r_0 A^{1/3}$. A small correction to $\asya$ from
surface compression \cite{mye69} is neglected here. Let us mention
that (\ref{eq4}) may be derived also from the notion of surface
symmetry energy \cite{ste05,dan03}.

The neutron skin thickness of nuclei is obtained as
\beq\label{eq2}
\sk=\sqrt{3/5} \lf[ t - e^2 Z / (70J) \ri] + \sw
\eeq
in the DM \cite{mye69,mye80}. The quantity $t$ gives the distance
between the neutron and proton mean surface locations:
\begin{eqnarray}\label{eq3}
 t &=& \frac{3 r_0}{2} \, \frac{J/Q}{1+x_A} \, (I-I_{\rm C}) 
\nonumber \\
   &=& \frac{2r_0}{3J} \, \lf[J-\asya\ri] A^{1/3} \, (I-I_{\rm C}) ,
\end{eqnarray}
where in the second line we have introduced the surface symmetry term
$a_{\rm ss}(A) = [J-\asya] A^{1/3}$ using Eq.\ (\ref{eq4}). The second
term in Eq.\ (\ref{eq2}) is due to Coulomb repulsion, and $\sw=
\sqrt{3/5}\lf[5 (b_n^2-b_p^2)/(2R)\ri]$ is a correction caused by an
eventual difference in the surface widths $b_n$ and $b_p$ of the
neutron and proton density profiles. 

\begin{figure}[t]
\includegraphics[width=0.95\columnwidth,angle=0,clip=true]{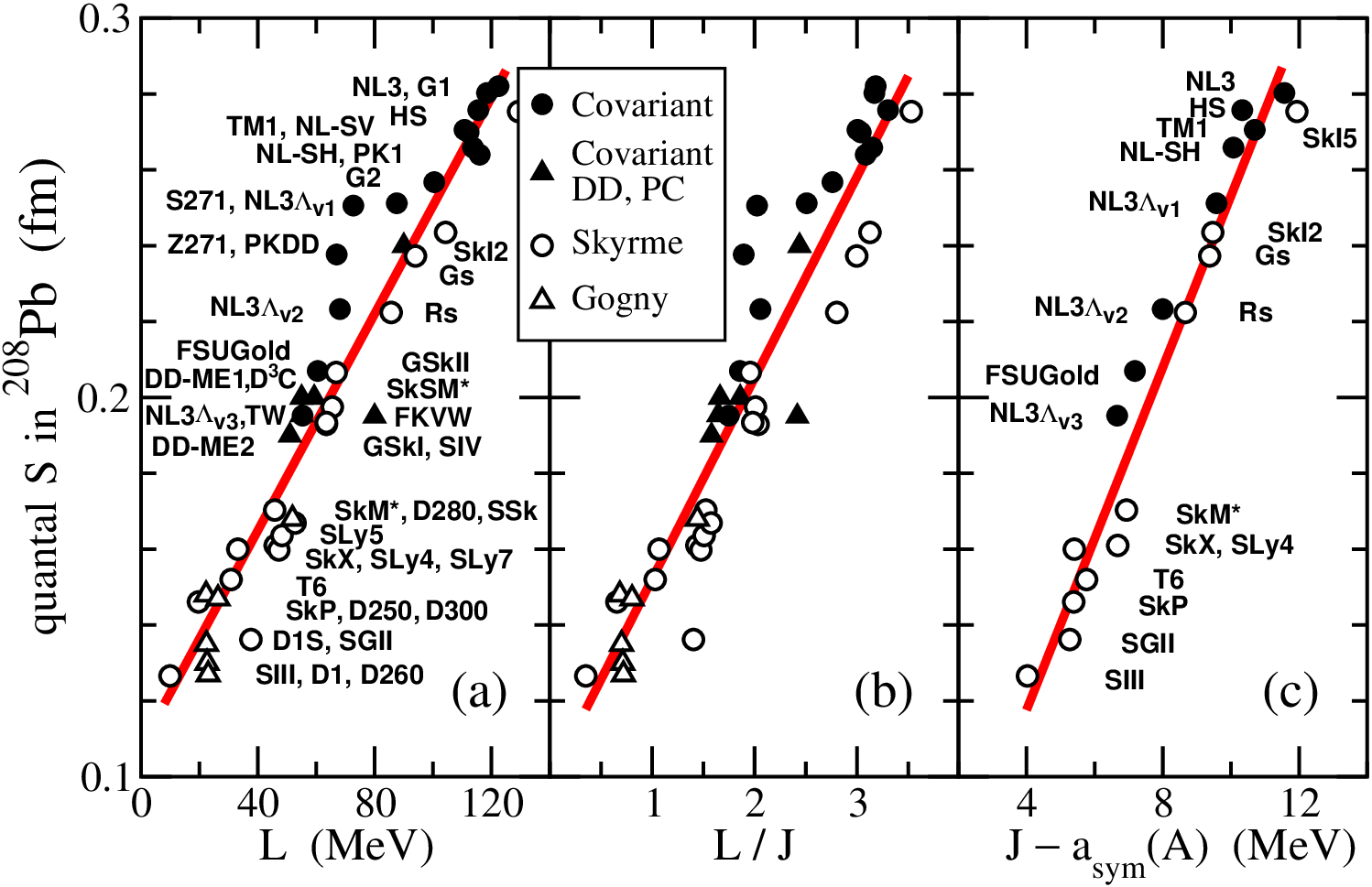}
\caption{(Color online) Correlation of the quantal selfconsistent
$\sk$ value in $^{208}$Pb with the slope of the symmetry energy $L$
(a), the ratio $L/J$ (b), and with $J-\asya$ (c), for various nuclear
models (DD and PC stand for density dependent and point coupling
models). From left to right, the correlation factors are
$r=0.961$, 0.945 and 0.970.}\end{figure}

We first illustrate the aforesaid correlation of $\sk$ of heavy nuclei
with $L$ in Fig.\ 1(a). It depicts the {\em quantal} self-consistent
value of $\sk$ in $^{208}$Pb against $L$ for multiple Skyrme, Gogny,
and covariant models of different nature
\cite{bro00,hor01,ste05,bar05,li05,li08,fur02,bra85,bla95}. In Fig.\
1(b) we show that a similar correlation exists with the ratio $L/J$,
which is proportional to $\gamma$ if a scaling $(\rho/\rho_0)^\gamma$
holds for $\csy(\rho)$. And in Fig.\ 1(c) we show that the close
dependence of $\sk$ on $J-\asya$ predicted by the DM is borne out in
the quantal $\sk$ value, using forces where we have computed $Q$ in
ASINM\@. It reassures one that the DM expression incorporates the
proper elements for the study. Many of the given nuclear interactions
are accurately fitted to experimental binding energies,
single-particle data, and charge radii of a variety of nuclei.
However, their isospin structure is not sufficiently firmed up as seen
e.g.\ in the differing predictions for $\sk({^{208}{\rm Pb}})$. There
is thus a need to deepen our knowledge of isospin-sensitive
observables like $\sk$ and of their constraints on $\csy(\rho)$.

\begin{table}\caption{Value of $J$, $\asya$ and density $\rho$ that
fulfils $\csy(\rho)=\asya$ for $A=208,116$ and 40, in various nuclear 
models. $J$ and $\asy$ are in MeV and $\rho$ is in fm$^{-3}$. Here
$\csy(\rho)$ was computed exactly as $\half\,\partial^2 e(\rho,\delta)/
\partial\delta^2 |_{\delta=0}\,$ from the EOS of the models.} 
\begin{ruledtabular}\begin{tabular}{lccccccccc}
&&\multicolumn{2}{c}{$A=208$} &&\multicolumn{2}{c}{$A=116$}
&&\multicolumn{2}{c}{$A=40$}\\
\cline{3-4} \cline{6-7} \cline{9-10} Model & $J$ &
$\asy$& $\rho$&& $\asy$& $\rho$&& $\asy$& $\rho$\\ \hline
NL3           &37.4&25.8&0.103&&24.2&0.096&&21.1&0.083\\
NL-SH         &36.1&26.0&0.105&&24.6&0.099&&21.3&0.086\\
FSUGold       &32.6&25.4&0.099&&24.2&0.092&&21.9&0.078\\
TF \cite{TF94}&32.6&24.2&0.094&&22.9&0.086&&20.3&0.071\\
SLy4          &32.0&25.3&0.100&&24.2&0.093&&22.0&0.079\\
SkX           &31.1&25.7&0.103&&24.8&0.096&&22.8&0.084\\
SkM*          &30.0&23.2&0.101&&22.0&0.094&&19.9&0.079\\
SIII          &28.2&24.1&0.093&&23.4&0.088&&21.8&0.078\\
SGII          &26.8&21.6&0.104&&20.7&0.098&&18.9&0.084
\end{tabular}\end{ruledtabular}\end{table}

We bring into notice a genuine relation between the symmetry energy
coefficients of the EOS and of nuclei: $\csy(\rho)$ equals $\asya$ of
a heavy nucleus like $^{208}$Pb at a density $\rho\approx0.1$
fm$^{-3}$. Indeed, the relation holds similarly down to medium mass
numbers, at lower $\rho$ values and a little more spread. Table I
exemplifies this fact with several nuclear models, where we show the
density fulfilling $\csy(\rho)=\asya$ for $A=208$, 116, and 40. We
find that this density can be parametrized as
 \beq\label{ansatz}
 \rho_A=\rho_0-\rho_0/(1+c A^{1/3}) 
 \eeq                       
with $c$ fixed by $\rho_{\,208}=0.1$ fm$^{-3}$ (which gives
$\rho_{\,116}\approx0.093$ fm$^{-3}$ and
$\rho_{\,40}\approx0.08$ fm$^{-3}$ for the models of Table I).

The relation ``$\csy(\rho)=\asya$'' can be very helpful to elucidate
other correlations of isospin observables with $\csy(\rho)$ and to
gain deeper insights into them. For example, it allows one to replace
$\asya$ in Eq.~(\ref{eq3}) for a heavy nucleus by $\csy(\rho)\simeq J
- L\epsilon + \half\Ksy\epsilon^2$ with $\epsilon$ computed at
$\rho\approx0.1 {\rm\,fm}^{-3}$ \cite{fn1}:
\beq\label{jqt}
t= \frac{2r_0}{3J} \, L \,
 \Big(1 - \epsilon \frac{\Ksy}{2L}\Big) \epsilon A^{1/3} 
 \big(I-I_{\rm C}\big) .
\eeq
The imprint of the density content of the symmetry energy on the
neutron skin appears now explicitly. The leading proportionality of
(\ref{jqt}) with $L$ explains the observed linearity of $\sk$ of a
heavy nucleus with $L$ in nuclear models \cite{bro00,li08,ste05}. The
correction with $\Ksy$ does not alter the situation as
$\epsilon\sim1/9$ is small. One can use Eq.\ (\ref{jqt}) in other mass
regions by calculating $\epsilon$ from $\rho_A$ of Eq.\
(\ref{ansatz}). We have checked numerically in multiple forces that
the results closely agree with Eq.\ (\ref{eq3}) for the $40\leq
A\leq238$ stable nuclei given in Fig.\ 2.

With the help of Eq.\ (\ref{jqt}) for $t$ (using $\rho_A$ to
compute~$\epsilon$), we next analyze constraints on the density
dependence of the symmetry energy by optimization of (\ref{eq2}) to
experimental $\sk$ data. We employ $\csy(\rho)= 31.6
(\rho/\rho_0)^\gamma$ MeV \cite{li05,li08,fam06,she07} and take as
experimental baseline the neutron skins measured in 26 antiprotonic
atoms \cite{trz01} (see Fig.\ 2). These data constitute the largest
set of uniformly measured neutron skins over the mass table till date.
With allowance for the error bars, they are fitted linearly by $\sk=
(0.9\pm0.15)I +(-0.03\pm0.02)$ fm \cite{trz01}. This systematics
renders comparisons of skin data with DM formulas, which by
construction average the microscopic shell effect, more meaningful
\cite{swi05}. We first set $b_n=b_p$ (i.e., $\sw=0$) as done in the
DM \cite{mye69,mye80,swi05} and in the analysis of data in Ref.\
\cite{dan03}. Following the above, we find $L=75\pm25$ MeV
($\gamma=0.79\pm0.25$). The range $\Delta L=25$ MeV stems from the
window of the linear averages of experiment. The $L$ value and its
uncertainty obtained from neutron skins with $\sw=0$ is thus quite
compatible with the quoted constraints from isospin diffusion and
isoscaling observables in HIC \cite{li05,li08,she07}. On the other
hand, the symmetry term of the incompressibility of the nuclear EOS
around equilibrium ($K=K_v + K_\tau\delta^2$) can be estimated using
information of the symmetry energy as $K_\tau\approx \Ksy-6L$
\cite{bar05,li05,li08}. The constraint $K_\tau=-500\pm50$ MeV is found
from isospin diffusion \cite{li05,li08}, whereas our study of neutron
skins leads to $K_\tau=-500^{+125}_{-100}$ MeV. A value
$K_\tau=-550\pm100$ MeV seems to be favored by the giant monopole
resonance (GMR) measured in Sn isotopes as is described in
\cite{gar07}. Even if the present analyses may not be called
definitive, significant consistency arises among the values extracted
for $L$ and $K_\tau$ from seemingly unrelated sets of data from
reactions, ground-states of nuclei, and collective excitations.

To assess the influence of the correction $\sw$ in (\ref{eq2}) we
compute the surface widths $b_n$ and $b_p$ in ASINM~\cite{cen98}. This
yields the $b_{n(p)}$ values of a finite nucleus if we relate the
asymmetry $\delta_0$ in the bulk of ASINM to $I$ by $\delta_0 (1+x_A)
= I+x_A I_{\rm C}$ \cite{mye80,bra85,cen98}. In doing so, we find that
Eq.\ (\ref{eq2}) reproduces trustingly $\sk$ (and its change with $I$)
of self-consistent Thomas-Fermi calculations of finite nuclei made
with the same nuclear force. Also, $\sw$ is very well fitted by
$\sw=\sigma_{\rm sw} I$. All slopes $\sigma_{\rm sw}$ of the forces of
Fig.\ 1(c) lie between $\sigma_{\rm sw}^{\rm min}= 0.15$ fm (SGII) and
$\sigma_{\rm sw}^{\rm max}=0.31$ fm (NL3). We then reanalyze the
experimental neutron skins including $\sw^{\rm min}$ and $\sw^{\rm
max}$ in Eq.\ (\ref{eq2}) to simulate the two conceivable extremes of
$\sw$ according to mean field models. The results are shown in Fig.~3.
Our above estimates of $L$ and $K_\tau$ could be shifted by up to
$-25$ and $+125$ MeV, respectively, by nonzero $\sw$. This is on the
soft side of the HIC \cite{li05,li08,she07} and GMR \cite{gar07}
analyses of the symmetry energy, but closer to the alluded predictions
from nucleon emission ratios \cite{fam06}, the GDR \cite{tri08}, and
nuclear binding systematics \cite{TF94}. One should mention that the
properties of $\csy(\rho)$ derived from terrestrial nuclei have
intimate connections to astrophysics \cite{hor01,ste05,lat07}. As an
example, we can estimate the transition density $\rho_t$ between the
crust and the core of a neutron star \cite{hor01,lat07} as
$\rho_t/\rho_0 \sim2/3 + (2/3)^\gamma \Ksy/2 K_v$, following the model
of Sect.\ 5.1 of Ref.\ \cite{lat07}. The constraints from neutron
skins hereby yield $\rho_t\sim 0.095\pm0.01$ fm$^{-3}$. This value
would not support the direct URCA process of cooling of a neutron star
that requires a higher $\rho_t$ \cite{hor01,lat07}. The result is in
accord with $\rho_t\sim0.096\, {\rm fm}^{-3}$ of the microscopic EOS
of Friedman and Pandharipande \cite{lor93}, as well as with
$\rho_t\sim 0.09\, {\rm fm}^{-3}$ predicted by a recent analysis of
pygmy dipole resonances in nuclei~\cite{kli07}.

\begin{figure}[t]
\includegraphics[width=0.90\columnwidth,angle=0,clip=true]{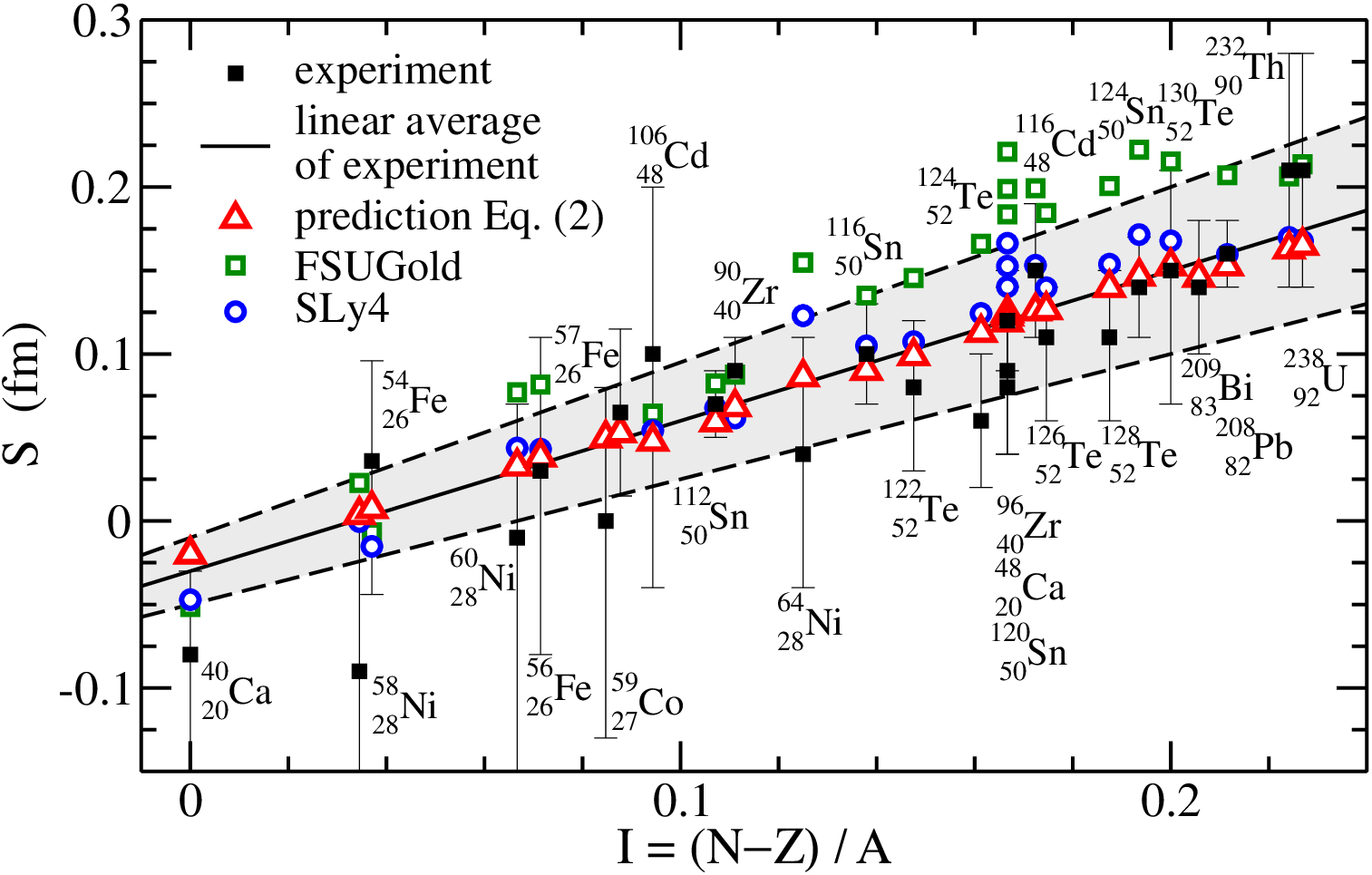}
\caption{(Color online) Comparison of the fit described in the text of
Eq.\ (\ref{eq2}) with the experimental neutron skins from
antiprotonic measurements and their linear average $\sk=(0.9\pm0.15)
I + (-0.03\pm0.02)$ fm \cite{trz01}. Results of the modern
Skyrme SLy4 and relativistic FSUGold forces are also
shown.}\end{figure}

\begin{figure}[t]
\includegraphics[width=0.90\columnwidth,angle=0,clip=true]{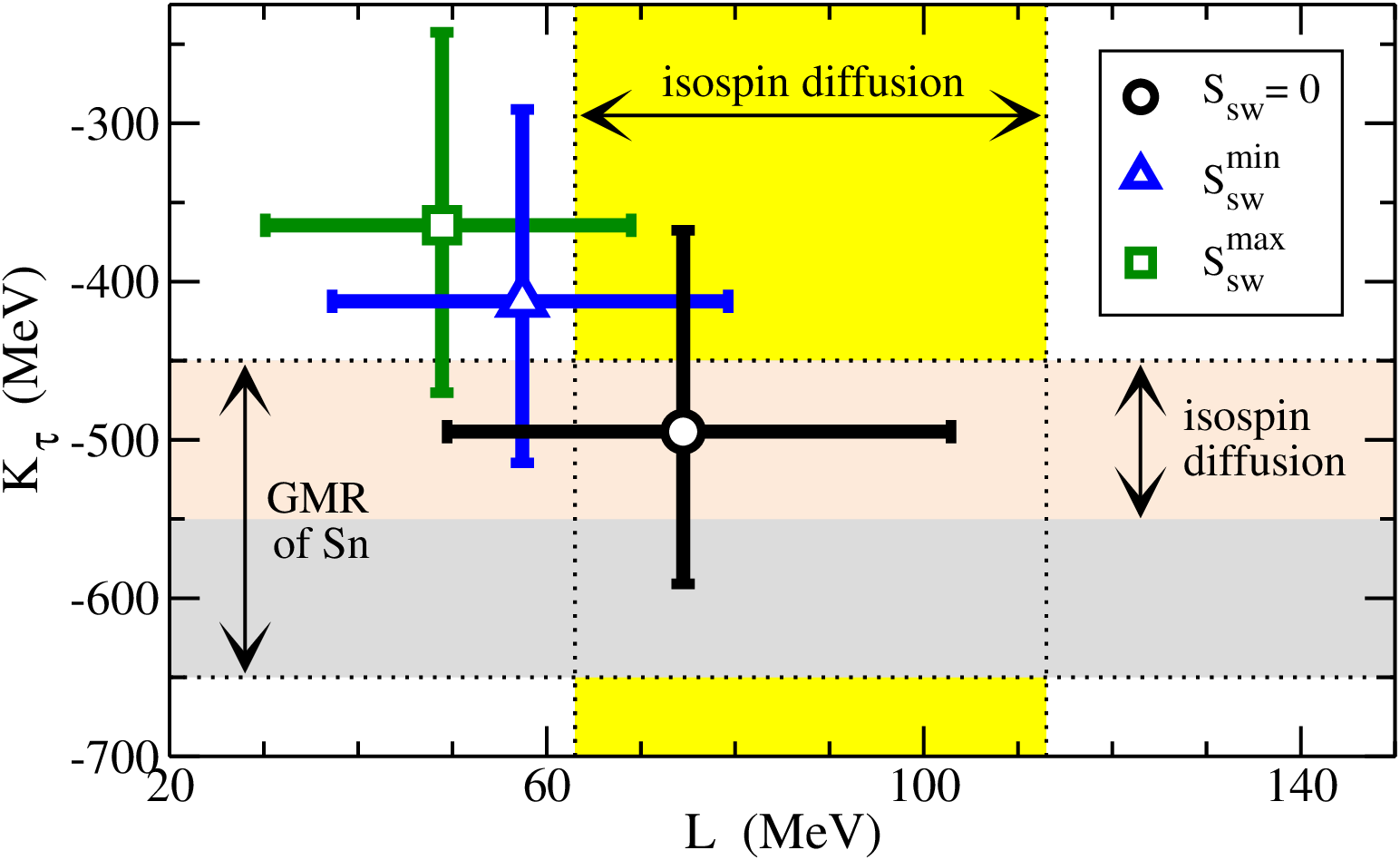}
\caption{(Color online) Constraints on $L$ and $K_\tau$ from neutron
skins and their dependence on the $\sw$ correction of Eq.\ (\ref{eq2}).
The crosses express the $L$ and $K_\tau$ ranges compatible with the
uncertainties in the skin data. The shaded regions depict the
constraints on $L$ and $K_\tau$ from isospin diffusion
\cite{li05,li08} and on $K_\tau$ as determined in \cite{gar07}
from the GMR of Sn isotopes.}\end{figure}

We would like to close with a brief comment regarding the GDR. As
mentioned, Ref.\ \cite{tri08} very interestingly constrains
$\csy(0.1)$ from the GDR of $^{208}$Pb. The analysis notes that the
mean excitation energy of the GDR depends on $g(A)= J/\{1+\frac{5}{3}
a_{\rm ss}(A) A^{-1/3} / J\}$ \cite{ste05,tri08} and shows numerically
that the values of $g(208)$ and $\csy(0.1)$ are correlated in Skyrme
forces. Inserting $a_{\rm ss}(A)$ given below Eq.\ (\ref{eq3}), one has
$g(A)= J/\{1+\frac{5}{3} [J-\asya]/J\}$. Immediately, the equivalence
$\asy(208)\approx\csy(0.1)$ explains why $g(208)$ has a dependence on
$\csy(0.1)$, gives it analytically, and validates it for any type of
mean field model \cite{fn2}. One could extend it to other $A$ values
through Eq.\ (\ref{ansatz}). In conclusion, the discussed
relation of $\csy(\rho)$ with $\asya$ can be much valuable to link
different problems depending upon $\asya$ of nuclei to the symmetry
properties of the EOS.

Summarizing, we have described a generic relation between the symmetry
energy in finite nuclei and in nuclear matter at subsaturation. It
plausibly encompasses other prime correlations of nuclear observables
with the density content of the symmetry energy. We take advantage of
this relation to explore constraints on $\csy(\rho)$ from neutron
skins measured in antiprotonic atoms \cite{trz01}. We discuss the $L$
and $K_\tau$ values that skins favor vis-\`a-vis most recent
observations from reactions and giant resonances. The difficult
experimental extraction of neutron skins limits their potential to
constrain $\csy(\rho)$. Interestingly, we learn that in spite of
present error bars in the data of \cite{trz01}, the size of the final
uncertainties in $L$ or $K_\tau$ is comparable to the other analyses.
This highlights the value of having skin data consistently measured
across the mass table, and calls for pursuing extended measurements of
neutron radii and skins with ``conventional'' hadronic probes.
Combined with a precision extraction of $R_n$ of $^{208}$Pb through
electroweak probes \cite{prex}, they would contribute to cast uniquely
tight constraints on $\csy(\rho)$.

Work supported by the Spanish Consolider-Ingenio 2010 Programme CPAN
CSD2007-00042 and grants Nos.\ FIS2008-01661 from MEC (Spain) and
FEDER, 2005SGR-00343 from Generalitat de Catalunya, and N202 179
31/3920 from MNiSW (Poland). X.R. acknowledges grant AP2005-4751 from
MEC (Spain).


\begin{thebibliography}{99}
\bibitem{fn1} The expression $J - L \epsilon + \half \Ksy \epsilon^2$
differs from the exact value of $\csy(\rho)$ by less than 1\% at
$\rho=0.1$ fm$^{-3}$ in all forces tested, thus being of sufficient
accuracy.
\bibitem{fn2} In practice this dependence becomes almost linear: using
for calculation the forces of Fig.\ 1(c) as well as the TF model of
\cite{TF94}, we find that $g(208)$ has a correlation factor
0.991 with $\asy(208)$ and 0.983 with $\csy(0.1)$.
\end{thebibliography}
\end{document}